# Low Frequency Electronic Noise in Single-Layer MoS$_2$ Transistors

*Vinod K. Sangwan,*[1,||] *Heather N. Arnold,*[1,||] *Deep Jariwala,*[1] *Tobin J. Marks,*[1,2] *Lincoln J. Lauhon,*[1] *and Mark C. Hersam*[1,2,*]

[1]Department of Materials Science and Engineering, Northwestern University, Evanston, Illinois 60208, USA.

[2]Department of Chemistry, Northwestern University, Evanston, Illinois 60208, USA.

*e-mail: m-hersam@northwestern.edu

ABSTRACT: Ubiquitous low frequency 1/*f* noise can be a limiting factor in the performance and application of nanoscale devices. Here, we quantitatively investigate low frequency electronic noise in single-layer transition metal dichalcogenide MoS$_2$ field-effect transistors. The measured 1/*f* noise can be explained by an empirical formulation of mobility fluctuations with the Hooge parameter ranging between 0.005 and 2.0 in vacuum (< 10$^{-5}$ Torr). The field-effect mobility decreased and the noise amplitude increased by an order of magnitude in ambient conditions, revealing the significant influence of atmospheric adsorbates on charge transport. In addition, single Lorentzian generation-recombination noise was observed to increase by an order of magnitude as the devices were cooled from 300 K to 6.5 K.

TEXT: Recently, ultrathin films of transition metal dichalcogenides (TMDCs) have attracted significant attention due to their unique electrical and optical properties.[1-4] In particular, single-layer $MoS_2$ is being heavily explored for low-power digital electronics,[5-7] light detection[8,9] and emission,[10] valley-polarization,[4] and chemical sensing applications.[11] However, inherent low frequency electronic noise (i.e., 1/$f$ noise or flicker noise) could limit the ultimate performance of $MoS_2$ for these applications. On the other hand, 1/$f$ noise may also be a useful tool for sensing technologies.[12,13] Although 1/$f$ noise is ubiquitous in solid-state electronics, it becomes even more pronounced in devices with reduced dimensions/size.[14-19] Consequently, the 'all-surface' structure of two-dimensional (2D) materials such as graphene and TMDCs make them extremely sensitive to random perturbations in the local environment.[1,20] Furthermore, unlike zero bandgap graphene, the emerging 2D semiconductors with a finite bandgap present a new platform to study low-frequency electronic noise. Despite extensive electrical characterization of bulk TMDCs[21] and more recently ultrathin forms of semiconducting TMDCs,[1] low frequency noise has not yet been quantitatively studied in these emerging van der Waals layered materials.

In this Letter, we analyze low frequency conductance fluctuations in high mobility (up to 65 $cm^2$/Vs at room temperature) single-layer $MoS_2$ (SL-$MoS_2$) field-effect transistors (FETs). Experimental data are analyzed using models that have previously been applied to 1/$f$ noise in Si metal-oxide-semiconductor FETs (MOSFETs)[18] and nanoscale transistors such as carbon nanotubes (CNTs)[22, 23] and graphene.[24, 25] We observe that $1/f$ noise in single-layer MoS$_2$ FETs follows the Hooge empirical law in the accumulation regime (*i.e.,* when the gate voltage ($V_g$) is larger than the threshold voltage ($V_{th}$)) with a Hooge parameter varying over the range of 0.005 to 2.0 in vacuum (< 10$^{-5}$ Torr). Furthermore, the noise amplitude scales linearly with the total number of carriers in devices fabricated on single MoS$_2$ flakes, confirming that $1/f$ noise is due to fluctuations in carrier mobility and not fluctuations in the number of carriers.[14, 26] In ambient conditions, the noise amplitude and Hooge parameter increase by an order of magnitude, highlighting the strong influence of atmospheric adsorbates on SL-MoS$_2$. The Hooge parameter also shows an inverse relationship with field-effect mobility ($\mu_{FET}$) in a manner similar to organic thin-film transistors[27] and graphene FETs.[28] Finally, generation-recombination (GR) noise[29-31] is observed in SL-MoS$_2$ FETs and increases by an order of magnitude as the devices are cooled from 300 K to 6.5 K.

Single-layer MoS$_2$ flakes were obtained *via* mechanical exfoliation on thermally oxidized (300 nm thick SiO$_2$) Si substrates. The single-layer thickness of the MoS$_2$ flakes was confirmed by Raman spectroscopy as discussed in Supporting Information S1. Two-probe FETs were fabricated using standard e-beam lithography and lift-off processes with Au electrodes (without an adhesion layer) to obtain quasi-ohmic contacts to MoS$_2$ (see the optical image in the inset of Fig. 1a).[32] Conductance fluctuations were measured using a low-noise current pre-amplifier and spectrum analyzer. Measurements were conducted in vacuum (< 10$^{-5}$ Torr) as well as in ambient conditions. Linear output characteristics (*I-V*) of a typical single-layer MoS$_2$ FET at drain biases $|V_d|$ < 0.5 V (Fig. 1a) suggest the absence of a large Schottky barrier at the contacts in vacuum. Transfer characteristics (drain current $I_d$ *versus* gate voltage $V_g$) of the same device (Fig. 1b) reveal n-type behavior with $\mu_{FET}$ = 34.1 cm$^2$/Vs and a current on/off ratio greater than 5 x 10$^5$ for $V_g$ = 60 V to –60 V (note that the off-current of ~10 pA is limited by the measurement setup) in agreement with recently reported[32] high mobility MoS$_2$ transistors.

Fig. 1c shows time-domain current fluctuations of the devices increasing with applied $V_g$. The 1/$f$ noise is often expressed using the Hooge empirical law:[14, 15]

$$S_I = \frac{AI^\gamma}{f^\beta} \qquad (1)$$

where $S_I$ is the current power spectral density, $I$ is the mean device current, $f$ is the frequency, and $A$ is the noise amplitude. The exponents, $\beta$ and $\gamma$, are ideally expected to be close to 1 and 2, respectively. The current noise spectral density ($S_I$) of a SL-MoS$_2$ device shows a 1/$f^\beta$ dependence with $\beta$ = 1.07 ± 0.01 up to a frequency of 8 kHz (Fig. 1d). Similarly, all 10 of the measured devices followed 1/$f^\beta$ behavior with $\beta$ = 1.0 ± 0.1 at room temperature. The exponent $\gamma$ = 2 suggests that 1/$f$ noise is an equilibrium phenomenon[17] and current fluctuations are caused by fluctuations in resistance as opposed to being driven by the applied current. All devices showed $\gamma$ = 2 ± 0.15 in vacuum (*e.g.*, $S_I$ scales as $I^{2.06\pm0.05}$ at $f$ = 10 Hz as shown in Supporting Information Fig. S2a). The constant $A$ is related to the total number of carriers ($N$) in the channel via $A = \frac{\alpha_H}{N}$, where $\alpha_H$ is the Hooge parameter. We obtain the noise amplitude $A$ by plotting the inverse noise power ($I^2/S_I$) as a function of frequency $f$ ($I^2/S_I$ = (1/$A$)$f$, Supporting Information Fig. S2b).

Historically, two different models have been developed to explain 1/$f$ noise in metal-oxide-semiconductor FETs (MOSFETs) based on fluctuations in carrier mobility (Hooge[14, 15, 18]) or fluctuations in carrier number (McWhorter[17, 18, 33]). In field-effect devices, the number of carriers $N$ can be modulated by the gate voltage. Here, we limit noise characterization to the linear regime ($V_d < 0.5$ V) under overdrive conditions $|V_g - V_{th}| > 0$ so that $N$ can be approximated as $N = (V_g - V_{th}).L.W.c_g/e$, where $c_g$ is the gate capacitance per unit area (11.2 nF/cm$^2$ for a 300 nm SiO$_2$ layer), $e$ is the electronic charge, and $L$ and $W$ are channel length and width, respectively. The current power spectrum follows 1/$f$ behavior closely ($\beta = 0.98 - 1.05$, Fig. 1e) in the full range of applied gate voltages ($V_g = 10 - 50$ V, $V_{th} = -10$ V), and 1/$A$ follows the transfer curve closely (Supporting Information Fig. S3) with $A$ in the range of $0.6 - 1.7 \times 10^{-6}$. Fig. 1e shows a linear relation between 1/$A$ and $|V_g - V_{th}|$, in contrast to the parabolic dependence of 1/$A$ on $V_g$ (1/$A \propto |V_g - V_{th}|^2$) expected for the carrier number fluctuation model.[14, 15, 22] Thus, the gate dependence of $A$ is consistent with the Hooge model for mobility fluctuation.

The Hooge parameter ($\alpha_H$) was obtained from $1/A = B|V_g - V_{th}|$, where $B$ is $(L.W.c_g)/(\alpha_H.e)$. For a total of 10 devices, $\alpha_H$ varied between $5.7 \times 10^{-3}$ and 1.95. The lowest $\alpha_H$ values are comparable to those in single carbon nanotube FETs ($9.3 \times 10^{-3} - 0.53$)[22, 34, 35] but are up to 10 times larger than those in single-layer graphene FETs ($4 \times 10^{-4} - 10^{-3}$)[24, 25] on similar oxide dielectrics. On the other hand, the highest values of $\alpha_H$ found here are comparable to those in disordered systems such as organic thin-film transistors (OTFTs).[27, 36] Therefore, it appears that the noise in SL-MoS$_2$ FETs is not only limited by traps in the underlying oxide dielectric, but also can be increased by additional surface contamination/adsorbates, thus suggesting that noise levels could be reduced in suspended geometries[34, 37, 38] and/or *via* surface passivation.[39] We note that SL-MoS$_2$ showed a larger device-to-device variability in the Hooge parameter compared to graphene. This variability could arise from a greater sensitivity of MoS$_2$ to variations in processing conditions in the absence of optimized cleaning protocols such as thermal annealing. As will be seen later, the devices fabricated and measured under identical conditions showed a more uniform noise level.

To further confirm the Hooge relation $A = \frac{\alpha_H}{N}$, $N$ was explicitly varied by changing the channel area for devices fabricated and measured under identical conditions. In particular, four MoS$_2$ FETs were fabricated on a single SL-MoS$_2$ flake (see Supporting Information Fig. S4 for an optical image of the flake). The three-fold symmetry of SL-MoS$_2$ results in triangle-shaped single crystal flakes (edge length ~ 22 µm).[40] This geometry enabled the fabrication of devices with variable channel areas by taking advantage of the naturally varying $W$ while keeping $L$ constant (see inset in Fig. 2b).[28] Since the noise characteristics of these devices were measured under identical gating, temperature, and vacuum conditions, the carrier number $N$ is expected to be proportional to the channel area. Fig. 2a shows $1/A$ as well as $I_d$ as a function of $V_g$ for the four devices numbered '1' to '4' in the inset of Fig. 2b. Again, $1/A$ *versus* $V_g$ data follow the transfer characteristics in the accumulation regime. As expected, $I_d$ also is proportional to $W$ (Fig. 2b) and yields an average $\mu_{FET}$ = 37.8 ± 2.2 cm$^2$/Vs. The area-normalized noise amplitude data of each device overlaps (Fig. 2b), validating the Hooge formalism for 1/$f$ noise in MoS$_2$ transistors.

Due to their large surface area to volume ratios, charge transport and 1/$f$ noise characteristics of nanomaterials are extremely sensitive to atmospheric adsorbates.[22, 41-43] In this case, the $\mu_{FET}$ of SL-MoS$_2$ is an order of magnitude lower in ambient than in vacuum (see Supporting Information Fig. S5),[32] and the threshold voltage increases by 20 – 40 V. Despite this threshold voltage shift, noise measurements could still be conducted at large overdrives where the $I_d - V_g$ curve is linear. A larger non-linearity in output characteristics was observed in all devices at $|V_d| < 0.5$ V (see Supporting Information Fig. S5a), suggesting an increased effective Schottky barrier at the metal-semiconductor contacts in ambient conditions. While the current noise power spectra maintained $1/f^\beta$ behavior with $\beta$ close to unity within ±0.1 (Fig. 3a) in all 8 of the devices measured in ambient, a subset of devices deviates from ideal $I^2$ dependence of $S_I$. In particular, these devices showed a $S_I \propto I^\gamma$ dependence with $\gamma = 1.6 – 2.1$ in ambient ($S_I \propto I^{1.76 \pm 0.04}$ for the device in Fig. 1; see Supporting Information Fig. S6a). This current dependence for $S_I$ is consistent with non-ohmic contacts (nonlinear I-V characteristics), and was previously reported in OTFTs,[36, 44] CNT FETs,[45] and conducting polymers.[46] The increased 'effective' Schottky barrier height in ambient has been previously explained by modulation of the contact metal work function by adsorbate-induced dipoles near the contacts.[47] Nevertheless, the devices in ambient still obey the Hooge mobility fluctuation model ($1/A \propto |V_g - V_{th}|$ in Fig. 3b) in accumulation. Fig. 3c shows 1 to 3 orders of magnitude increase in $\alpha_H$ for SL-MoS$_2$ from vacuum to ambient with an inverse correlation between $\alpha_H$ and $\mu_{FET}$, in agreement with previous studies in percolating OTFTs,[27, 48] polymer transistors,[49] and graphene FETs.[28] Note that the effect of ambient conditions on $1/f$ noise in MoS$_2$ transistors stands in stark contrast to CNT FETs that exhibit up to 3-times reduced noise in ambient due to increased carrier concentration and thus increased conductance *via* ambient doping.[22]

Finally, a temperature-dependent study of current fluctuations in SL-MoS$_2$ transistors was conducted. Our high quality SL-MoS$_2$ FETs showed band-like transport with $\mu_{FET}$ increasing up to 2.5 times from 300 K to 6.5 K with the highest mobility of 146.7 cm$^2$/Vs at 6.5 K (Supporting Fig. S7).[32] The MoS$_2$ low frequency noise at low temperatures is adequately represented by a superposition of $1/f$ noise and one Lorentzian.[29, 50] The emergence of a single

Lorentzian in the noise spectra suggests generation-recombination (GR) noise that originates from fluctuations in the number of free carriers involving random transitions between states of different energy bands.[51-53] Fig. 4 shows the noise spectral density of a device at 6.5 K that was fit to:

$$S_I = \frac{AI^2}{f} + \frac{BI^2}{1+\left(f/f_0\right)^2} \quad (2)$$

where $A$ and $B$ are constants, and $f_0$ is the characteristic frequency of the generation-recombination process. The relative contribution of GR noise (*i.e.*, B/A ratio) increased by an order of magnitude from 300 K to 6.5 K (1.8 x $10^{-3}$ at 300 K to 2.1 x $10^{-2}$ at 6.5 K, see Supporting Information Fig. S8). Note that some devices show a shoulder in the noise spectral density even at room temperature, which suggests a larger GR noise contribution in those cases ($B/A \sim 10^{-3}$, Fig. S8). GR processes in the case of a single two-level fluctuator have been shown to generate random telegraph signals in individual CNT devices[54] and small channel MOSFETs.[55] Time-domain measurements on the present $MoS_2$ devices, however, do not reveal random telegraph features.

In conclusion, we have performed an extensive study of low-frequency electronic noise in high-quality unencapsulated single layer $MoS_2$ FETs. Carrier density (*via* gate voltage) and carrier number (*via* channel area) dependent studies revealed the Hooge mobility fluctuation model as the dominant source of $1/f$ noise in $MoS_2$ at room temperature. The extracted Hooge parameter ranges over two orders of magnitude (0.005 – 2.0) and increases by more than an order of magnitude in ambient conditions, suggesting a high sensitivity of SL-$MoS_2$ to adsorbates. The lowest values of the Hooge parameter are comparable to other "all-surface" nanomaterials such as CNTs on oxide gate dielectrics, which implicate dielectric quality in determining the 1/$f$ noise level. Additionally, the observation of low frequency generation-recombination noise at low temperature could be due to traps in the underlying $SiO_2$ substrate or midgap states in SL-MoS$_2$, presenting a unique diagnostic tool for trapping processes and materials purity analysis in ultrathin semiconductors.[56] Finally, these noise metrics are expected to provide useful guidelines for researchers as they develop high-performance electronic and sensing devices based on emerging single-layer transition metal dichalcogenides.

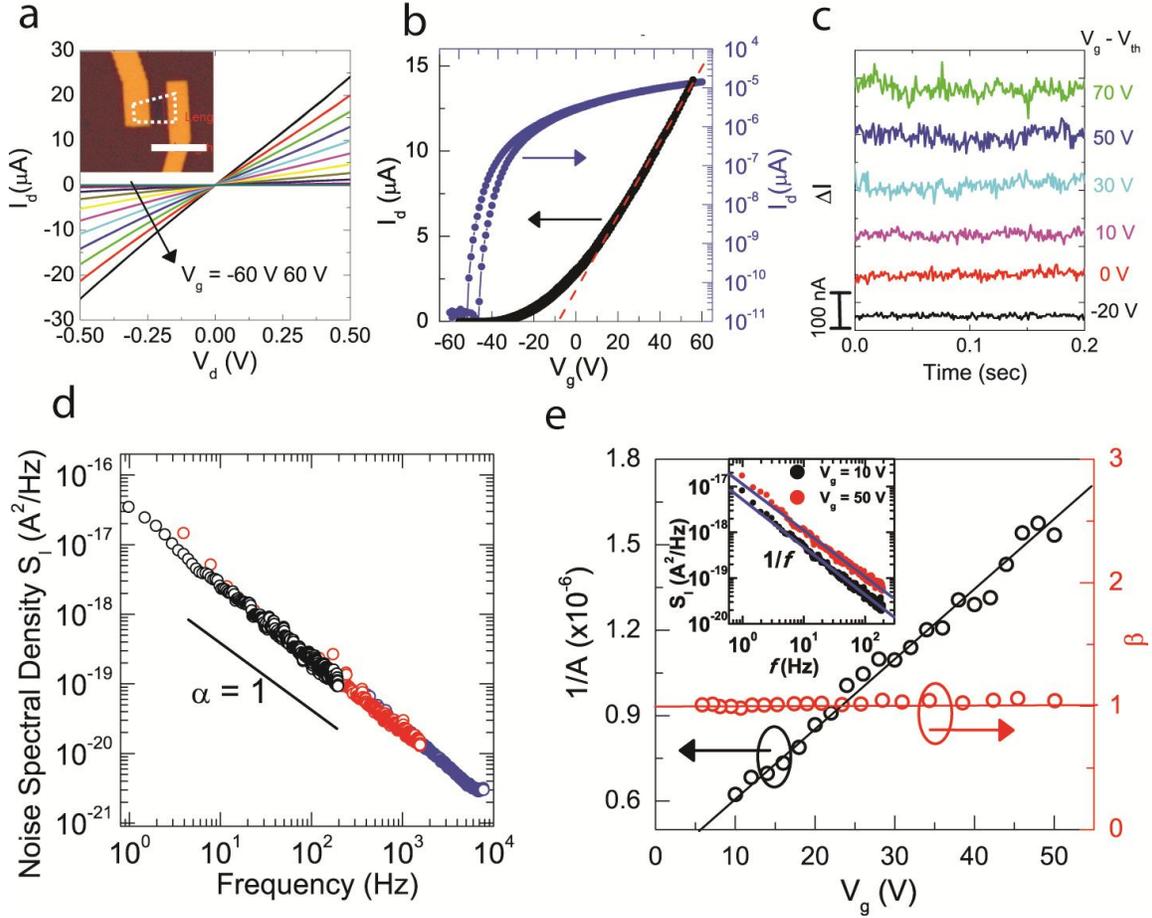

**Figure 1.** a) Output characteristics of a SL-MoS$_2$ field-effect transistor at 9 x 10$^{-6}$ Torr for gate bias ($V_g$) ranging from −60 V to 60 V in steps of 10 V. The inset shows an optical image of the device ($L$ = 1.71 µm, $W$ = 3.32 µm) where the SL-MoS$_2$ flake is outlined by a white dashed line. The white scale bar corresponds to 5 µm. b) Transfer characteristics of the same device at a drain bias $V_d$ = 0.3 V in both linear and log-linear plots. The red dashed lines show the threshold voltage $V_{th}$ = −10 V. c) Time domain current fluctuations at overdrive ($V_g - V_{th}$) ranging from −20 V to 70 V. d) Noise spectral density ($S_I$) as a function of frequency at $V_g$ = 20 V and $V_d$ = 0.2 V showing 1/$f^\beta$ behavior with $\beta$ = 1.07 ± 0.01. The black line shows ideal 1/$f$ behavior. e) Inverse noise amplitude 1/$A$ (left axis) and exponent $\beta$ (right axis) as a function of gate voltage ($V_g$) at $V_d$ = 0.1 V in vacuum (9 x 10$^{-6}$ Torr). The black line shows a linear fit ($r^2$ > 0.98) to the $1/A$ data that is used to extract the Hooge parameter. The inset shows the noise spectral density ($S_I$) versus frequency at two extreme values of $V_g$ = 10 V and 50 V ($V_{th}$ = –10 V). Blue lines are least-square fits to extract $\beta = 1.05 \pm 0.01$ ($V_g$ = 10 V) and $\beta = 0.98 \pm 0.01$ ($V_g$ = 50 V).

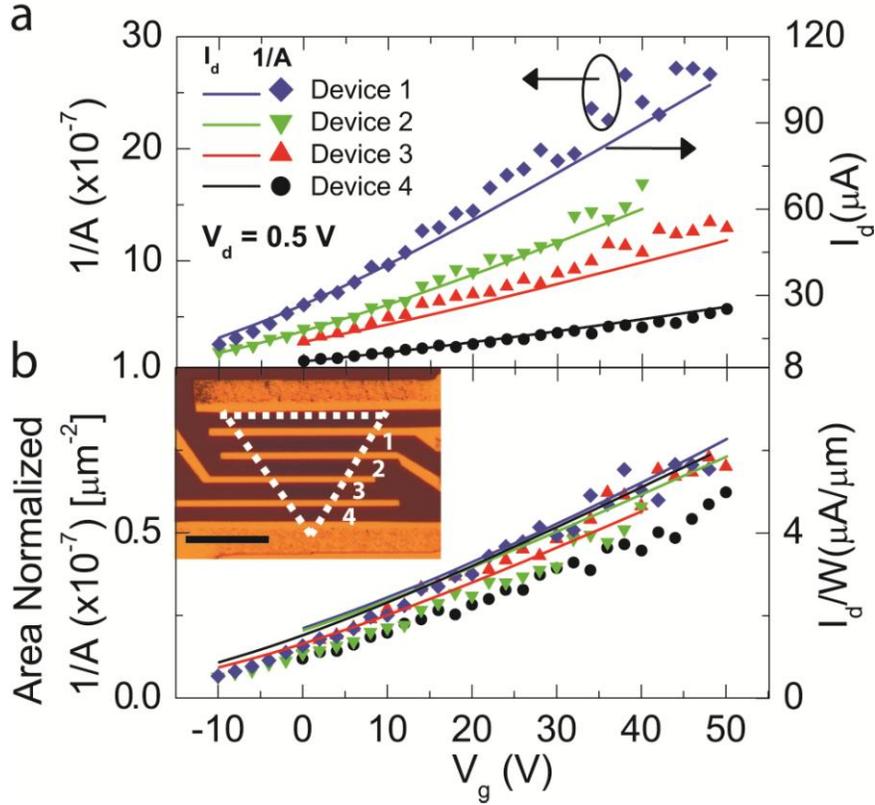

**Figure 2.** a) Inverse noise amplitude $1/A$ (left axis, symbols) and drain current $I_d$ (right axis, lines) *versus* $V_g$ for four different field-effect transistors (devices 1 – 4) fabricated on a single crystalline flake of SL-MoS$_2$. The legend is the same for parts (a) and (b). b) Area normalized $1/A$ and channel width normalized $I_d$ are plotted against $V_g$. The inset shows an optical micrograph of four devices (1 – 4) fabricated on a SL-MoS$_2$ flake outlined by the white dashed line. The black scale bar corresponds to 10 μm. Channel widths are calculated as the mean of the two parallel sides of the trapezoidal device channels.

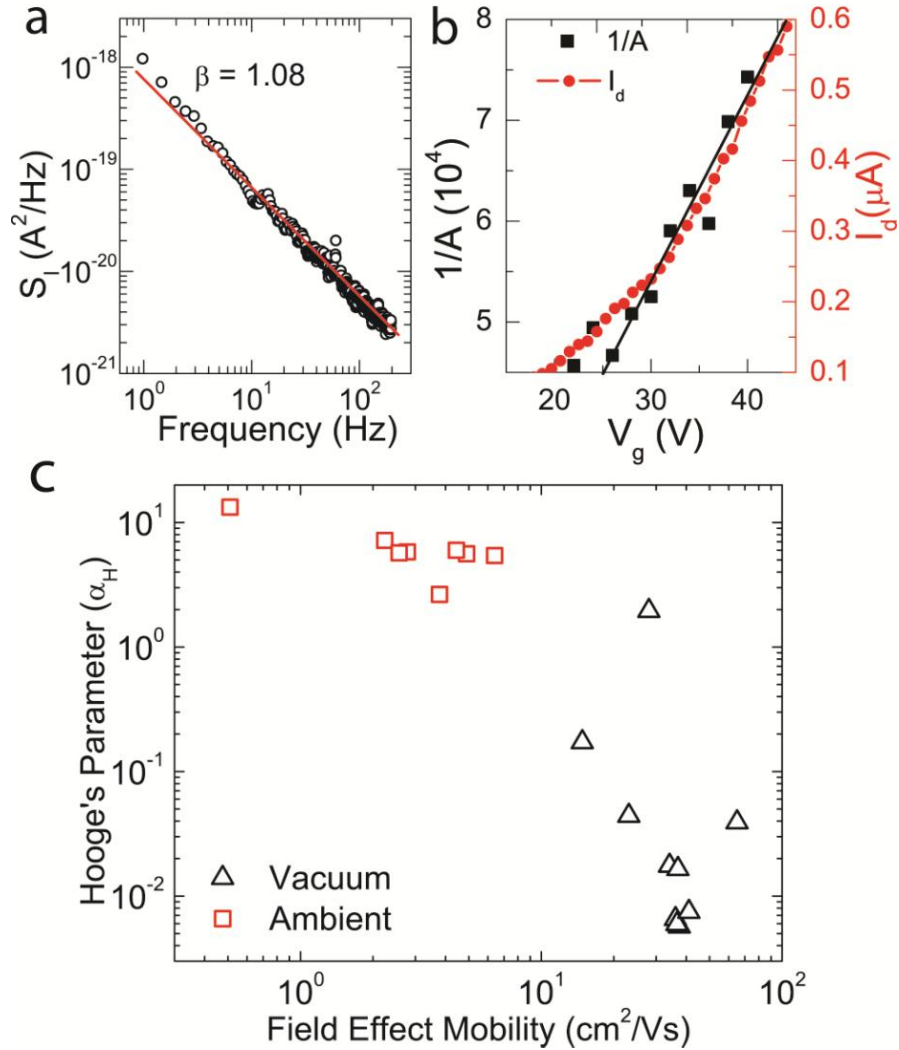

**Figure 3.** a) Noise spectral density ($S_I$) *versus* frequency in ambient conditions of the same device as Fig. 1 showing $1/f^\beta$ with $\beta = 1.08 \pm 0.01$. b) Inverse noise amplitude $1/A$ and drain current $I_d$ *versus* gate voltage $V_g$ at drain voltage $V_d = 0.3$ V in ambient conditions. The black line is a linear fit to $1/A$ in positive overdrive. c) The Hooge parameter $\alpha_H$ is plotted as a function of field-effect mobility for all the devices under vacuum as well as in ambient conditions.

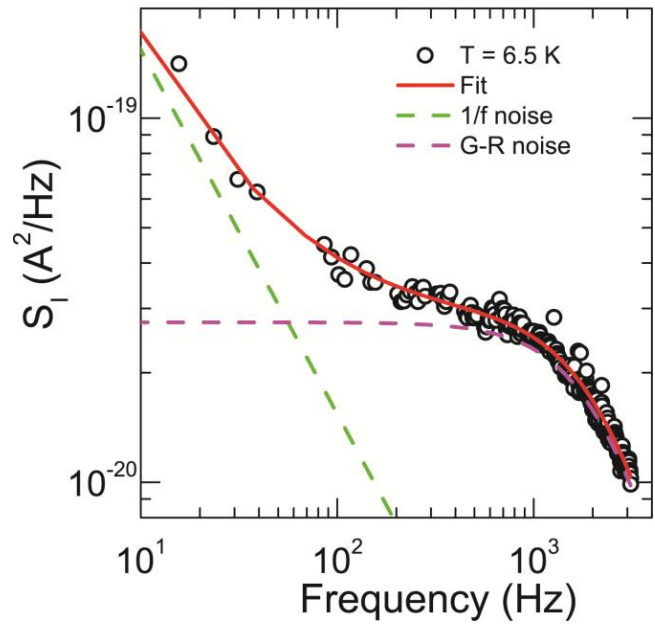

**Figure 4.** Noise spectral density of a device at 6.5 K as a function of frequency for $V_g$ = 45 V and $V_d$ = 0.3 V. Peaks at 60 Hz and harmonics are removed. Dashed lines show components of 1/$f$ noise and generation-recombination (GR) noise extracted by fitting the data to equation 2 (red line, $r^2$ > 0.98) with $f_0$ = 2317 Hz and $B/A$ ratio of 2.1 x $10^{-2}$.

## ASSOCIATED CONTENT

**Supporting Information**. Raman spectroscopy of single-layer $MoS_2$ flakes; 1/f noise data in vacuum and ambient conditions; variable temperature transport and noise data. This material is available free of charge via the Internet at http://pubs.acs.org.

## AUTHOR INFORMATION

**Corresponding Author**

Mark Hersam, m-hersam@northwestern.edu

**Author Contributions**

|| These authors contributed equally to the manuscript.

**Notes**

The authors declare no competing financial interest.

## ACKNOWLEDGMENT

This research was supported by the Materials Research Science and Engineering Center (MRSEC) of Northwestern University (NSF DMR-1121262). H.N.A. acknowledges support from a NASA Space Technology Research Fellowship. The authors acknowledge assistance from B. Myers and I. S. Kim with electron beam lithography and Raman spectroscopy, respectively. This research utilized the NUANCE Center at Northwestern University, which is supported by NSF-NSEC, NSF-MRSEC, Keck Foundation, and the State of Illinois.

**TOC Image:**

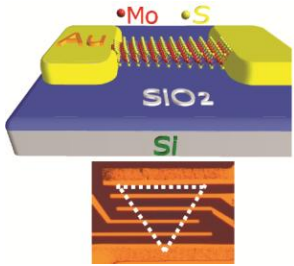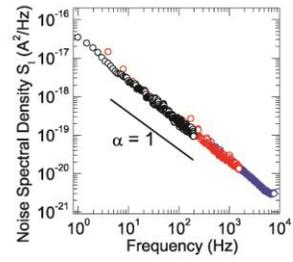